\documentclass[twocolumn,trackchanges]{aastex62} 
\usepackage[normalem]{ulem}
\usepackage{lineno}

\shorttitle{emulator for galaxy power spectrum}

\shortauthors{Wang et al.}

\def\ie{{\frenchspacing\it i.e.}}

\def\etc{{\frenchspacing\it etc.}}

\def\be{\begin{equation}}
\def\ee{\end{equation}}
\def\ba{\begin{eqnarray}}
\def\ea{\end{eqnarray}}

\begin{document}

\title{Emulating power spectra for pre- and post-reconstructed galaxy samples}

\correspondingauthor{Yuting Wang}
\email{ytwang@nao.cas.cn}

\author[0000-0001-7756-8479]{Yuting Wang}
\affiliation{National Astronomical Observatories, Chinese Academy of Sciences, Beijing, 100101, P.R.China}
\affiliation{Institute for Frontiers in Astronomy and Astrophysics, Beijing Normal University, Beijing 102206, China}

\author[0000-0002-7284-7265]{Ruiyang Zhao}
\affiliation{National Astronomical Observatories, Chinese Academy of Sciences, Beijing, 100101, P.R.China}
\affiliation{School of Astronomy and Space Science, University of Chinese Academy of Sciences, Beijing 100049, P.R.China}
\affiliation{Institute of Cosmology and Gravitation, University of Portsmouth, Dennis Sciama Building, Portsmouth PO1 3FX, United Kingdom}

\author[0000-0001-7984-5476]{Zhongxu Zhai}
\affiliation{Department of Astronomy, School of Physics and Astronomy, Shanghai Jiao Tong University, Shanghai 200240, China}
\affiliation{Shanghai Key Laboratory for Particle Physics and Cosmology, Shanghai 200240, China}
\affiliation{Waterloo Centre for Astrophysics, University of Waterloo, 200 University Ave W, Waterloo, ON, N2L 3G1, Canada}
\affiliation{Department of Physics and Astronomy, University of Waterloo, 200 University Ave W, Waterloo, ON, N2L 3G1, Canada}

\author{Kazuya Koyama}
\affiliation{Institute of Cosmology and Gravitation, University of Portsmouth, Dennis Sciama Building, Portsmouth PO1 3FX, United Kingdom}

\author{Will J. Percival}
\affiliation{Waterloo Centre for Astrophysics, University of Waterloo, 200 University Ave W, Waterloo, ON, N2L 3G1, Canada}
\affiliation{Department of Physics and Astronomy, University of Waterloo, 200 University Ave W, Waterloo, ON, N2L 3G1, Canada}
\affiliation{Perimeter Institute for Theoretical Physics, 31 Caroline St. North, Waterloo, ON, N2L 2Y5, Canada}

\author[0000-0003-4936-8247]{Hong Guo}
\affiliation{Shanghai Astronomical Observatory, Chinese Academy of Sciences, Shanghai 200030, China}

\author[0000-0002-0701-1410]{Yin Li}
\affiliation{Department of Mathematics and Theory, Peng Cheng Laboratory, Shenzhen, Guangdong 518066, China}

\author[0000-0003-4726-6714]{Gong-Bo Zhao}
\affiliation{National Astronomical Observatories, Chinese Academy of Sciences, Beijing, 100101, P.R.China}
\affiliation{School of Astronomy and Space Science, University of Chinese Academy of Sciences, Beijing 100049, P.R.China}
\affiliation{Institute for Frontiers in Astronomy and Astrophysics, Beijing Normal University, Beijing 102206, China}

\author{Takahiro Nishimichi}
\affiliation{Department of Astrophysics and Atmospheric Sciences, Faculty of Science,
Kyoto Sangyo University, Motoyama, Kamigamo, Kitaku, Kyoto 603-8555, Japan}
\affiliation{Center for Gravitational Physics and Quantum Information, Yukawa Institute for Theoretical Physics, 
Kyoto University, Kyoto 606-8502, Japan}
\affiliation{Kavli Institute for the Physics and Mathematics of the Universe (WPI), The University of Tokyo Institutes for Advanced Study (UTIAS), The University of Tokyo, Chiba 277-8583, Japan}

\author{H\'ector Gil-Mar\'{\i}n} 

\affiliation{ ICC, University of Barcelona, IEEC-UB, Mart\'{\i} i Franqu\`es, 1, E-08028 Barcelona, Spain}

\author{Yonghao Feng}
\affiliation{National Astronomical Observatories, Chinese Academy of Sciences, Beijing, 100101, P.R.China}
\affiliation{School of Astronomy and Space Science, University of Chinese Academy of Sciences, Beijing 100049, P.R.China}

\author{Hanyu Zhang}
\affiliation{Waterloo Centre for Astrophysics, University of Waterloo, 200 University Ave W, Waterloo, ON, N2L 3G1, Canada}
\affiliation{Department of Physics and Astronomy, University of Waterloo, 200 University Ave W, Waterloo, ON, N2L 3G1, Canada}

\author{Yi Wu}
\affiliation{National Astronomical Observatories, Chinese Academy of Sciences, Beijing, 100101, P.R.China}
\affiliation{School of Astronomy and Space Science, University of Chinese Academy of Sciences, Beijing 100049, P.R.China}

\begin{abstract}
The small-scale linear information in galaxy samples typically lost during non-linear growth can be restored to a certain level by the density field reconstruction, which has been demonstrated for improving the precision of the baryon acoustic oscillations (BAO) measurements. As proposed in the literature, a joint analysis of the power spectrum before and after the reconstruction enables an efficient extraction of information carried by high-order statistics. However, the statistics of the post-reconstruction density field are difficult to model. In this work, we circumvent this issue by developing an accurate emulator for the pre-reconstructed, post-reconstructed, and cross power spectra ($P_{\rm pre}$, $P_{\rm post}$, $P_{\rm cross}$) up to $k=0.5~h~{\rm Mpc^{-1}}$ based on the \textsc{Dark Quest} N-body simulations. The accuracy of the emulator is at percent level, namely, the error of the emulated monopole and quadrupole of the power spectra is less than $1\%$ and $10\%$ of the ground truth, respectively. A fit to an example power spectra using the emulator shows that the constraints on cosmological parameters get largely improved using $P_{\rm pre}$+$P_{\rm post}$+$P_{\rm cross}$ with $k_{\rm max}=0.25~h~{\rm Mpc^{-1}}$, compared to that derived from $P_{\rm pre}$ alone, namely, the constraints on ($\Omega_m$, $H_0$, $\sigma_8$) are tightened by $\sim41 \%-55\%$, and the uncertainties of the derived BAO and RSD parameters ($\alpha_{\perp}$, $\alpha_{||}$, $f\sigma_8$) shrink by $\sim 28\%-54\%$, respectively. This highlights the complementarity among $P_{\rm pre}$, $P_{\rm post}$ and $P_{\rm cross}$, which demonstrates the efficiency and practicability of a joint $P_{\rm pre}$, $P_{\rm post}$ and $P_{\rm cross}$ analysis for cosmological implications.

\end{abstract}

\section{Introduction} \label{sec:intro}
Wide-area spectroscopic surveys are fundamental tools for cosmological studies since they enable us to probe the Universe both geometrically and dynamically. In particular, the observed baryon acoustic oscillations (BAO) and redshift-space distortions (RSD), which are specific three-dimensional clustering patterns of galaxies, can be used to reconstruct the cosmic expansion history and the growth rate of the cosmic structure. Over the last few decades, massive spectroscopic surveys, including the Sloan Digital Sky Survey (SDSS) \citep{SDSS}, the Two-Degree-Field Galaxy Redshift Survey (2dFGRS) \citep{2dFGRS}, WiggleZ \citep{WiggleZ}, the SDSS-III Baryon Oscillation Spectroscopic Survey (BOSS) \citep{Dawson2013}, and the SDSS-IV extended Baryon Oscillation Spectroscopic Survey (eBOSS) \citep{Dawson2016} have proven to be a powerful probe for cosmology \citep{Peacock:2001gs,Eisenstein2005,Cole:2005sx,Percival2006,Blake2011,Alam:2016hwk,eBOSS:2020yzd}.

In Fourier space, the BAO feature manifests itself as a set of wiggles in the power spectrum, which can be used as a standard ruler to measure the cosmic expansion history. Unfortunately, the BAO feature is generally blurred by the nonlinear evolution of the cosmic structure, reducing its strength as a cosmic probe. To sharpen the BAO feature, the reconstruction scheme was proposed \citep{recon07}, which effectively restores the linearity of the density field to a certain extent by partially undoing the nonlinear structure evolution. This process brings the high-order information dominated by the $3$ -pt and $4$ -pt statistics back to $2$ -pt statistics \citep{Schmittfull:2015mja}, such that it is not only useful for boosting the BAO signal, but also helpful for a general full-shape analysis of the power spectrum \citep{Hikage2020}.

Recently, a novel method was proposed \citep{Wang:2022nlx} to extract information carried by high-order statistics from a joint analysis of the power spectrum of the pre-reconstructed density field ($P_{\rm pre}$), the post-reconstructed field ($P_{\rm post}$), and the cross-power spectrum between pre- and post-reconstructed fields ($P_{\rm cross}$). Their analysis, based on the Fisher matrix method, showed that a joint analysis using $P_{\rm pre}$, $P_{\rm post}$ and $P_{\rm cross}$ can tighten the constraints on the cosmological parameters compared to that using $P_{\rm post}$ alone, as part of the information from $3$ -pt and $4$ -pt of the density field can be efficiently extracted \citep{Wang:2022nlx}. 

In order to exploit the information content from the galaxy clustering, an accurate model for the statistics of the density field before and after the reconstruction is required. Traditional methods for the model building rely on the perturbation theory (PT). For $P_{\rm pre}$, PT-based models can work up to scales of $k=0.2$ or $0.25~h~{\rm Mpc^{-1}}$, depending on the effective redshift of the galaxy sample \citep{TNS,EFTLSS,eTNS,DAmico:2019fhj,Ivanov:2019pdj,Chen:2021wdi}. However, it is much more challenging to build PT-based models that can work on the same scales for $P_{\rm post}$ or $P_{\rm cross}$, due to complexities brought in by the reconstruction process \citep{Hikage:2019ihj}. One alternative to building PT-based models is to develop simulation-based models, \ie, the emulators, which have been extensively studied and developed for statistics for the pre-reconstructed density fields \citep{Zhai:2018plk,Wibking:2017slg,Kobayashi:2020zsw,Yuan:2022jqf,Winther:2019mus,Donald-McCann:2021nxc,Cuesta-Lazaro:2022dgr,Kwan:2023yph,Cuesta-Lazaro:2023gbv}.

In this work, we develop an emulator for $P_{\rm pre}, P_{\rm post}$ and $P_{\rm cross}$ up to $k=0.5~h~{\rm Mpc^{-1}}$, which is trained using the \textsc{Dark Quest} simulations \citep{Nishimichi:2018etk} and an halo occupation distribution (HOD) model \citep{Zheng07}. Our emulator is then validated using simulations that are not used for the training. Using our emulator, we perform a likelihood analysis using the monopole and quadrupole of galaxy power spectra up to the scale of $k =0.25~h~{\rm Mpc^{-1}}$, and find a significant information gain by a joint $\{ P_{\rm pre}, P_{\rm post}, P_{\rm cross}\}$ analysis, compared to using $P_{\rm pre}$ alone.

This paper is structured as follows: the next section is a description of the simulations and galaxy mocks used for the training and validation, and Sec. \ref{sec:emu} presents the details for creating the emulator. In Sec. \ref{sec:fit} we perform a likelihood analysis using various types of power spectra and show the main result of this work, before conclude in Sec. \ref{sec:conclusion}.

\section{The \textsc{Dark Quest} simulations and Galaxy mocks} \label{sec:sim}
The \textsc{Dark Quest} simulations that we use to develop our emulator are a suite of $N$-body simulations with $2048^3$ dark matter particles in $2~h^{-1} {\rm Gpc}$ side-length box \citep{Nishimichi:2018etk}. The emulator is built using a single Dark Quest snapshot at $z = 0.549$. The cosmologies used in the \textsc{Dark Quest} simulations cover the $100$ spatially flat $w$CDM models\footnote{These $100$ cosmological simulations are generated using different random number seeds \citep{Nishimichi:2018etk}.} with six variable parameters and one spatially flat $\Lambda$CDM model with the best-fit value of Planck 2015 \citep{Planck:2015fie} presented in Table \ref{tab:para}, where $\omega_b \equiv \Omega_b h^2$ and $\omega_c \equiv \Omega_c h^2$ are the physical density parameters of baryon and cold dark matter, respectively. $\Omega_{de}$ is the dimensionless dark energy density parameter. $A_s$ and $n_s$ are the amplitude and slope of the primordial power spectrum, respectively. $w$ is the equation of state parameter of dark energy. In addition, the total neutrino mass is fixed to $\sum m_\nu = 0.06\,\mathrm{eV}$. The effect of massive neutrino was included in simulations at the level of linear transfer function.  Cosmological parameters are sampled over the parameter range presented in Table \ref{tab:para} using optimal maximum distance sliced Latin hypercube designs \citep{Latin} so that parameter samplings can cover the parameter space as uniformly as possible \citep{Nishimichi:2018etk}. We have $15$ realizations for the fiducial $\Lambda$CDM cosmology.

The halos were identified using the phase-space temporal friends-of-friends halo finder, \textsc{Rockstar} \citep{rockstar}. The center of each halo is given as the center-of-mass location of a subset of member particles in the inner part of that halo, \ie, ``core particles", and the velocity of each halo is defined as the center-of-mass velocity of the core particles. $M_{\rm 200\mathrm{m}} = (4\pi/3)200\bar{\rho}_{\rm m0} R_{200\mathrm{m}}^2$ is used as the halo mass definition in \textsc{Dark Quest}, where $R_{200\mathrm{m}}$ is the spherical halo boundary radius within which the mean mass density is $200$ times the mean mass density today $\bar{\rho}_{\rm m0}$. The direct outputs of the \textsc{Rockstar} contain both distinct ``host'' halos and substructures. For the subsequent analyses, we remove substructures, which are found within the $R_{200\mathrm{m}}$ of a more massive nearby halo.

\begin{table}[!t]
\centering
\footnotesize
\begin{tabular}{ccc}
\hline\hline
Parameter  &  Fiducial value & Sampling range     \\\hline
$\omega_b$  &$0.02225$ &$[0.0211375,0.0233625]$\\
$\omega_c$  &$0.1198$ &$[0.10782,0.13178]$\\
$\Omega_{de}$  &$0.6844$ &$[0.54752,0.82128]$\\
$\ln(10^{10}A_s)$  &$3.094$ &$[2.4752,3.7128]$\\
$n_s$  &$0.9645$ &$[0.916275,1.012725]$\\
$w$  &$-1$ &$[-1.2,-0.8]$\\\hline
$\sigma_{{\rm log}M}$ & $0.596$ & $[0.05,1.2]$ \\
$M_0/M_1$ & $0.1194$ & $[0.0,0.4]$ \\
$\alpha$ & $1.0127$ & $[0.2,1.5]$ \\
$M_1/M_{{\rm min}}$ & $8.1283$ & $[5,15]$ \\ 
\hline\hline
\end{tabular}
\caption{Cosmological and HOD parameters used in our emulator. The fiducial cosmological values are from Planck 2015 \citep{Planck:2015fie}. We take the fiducial HOD parameters based on the fitting to CMASS galaxy sample \citep{Manera2013}. The sampling ranges represent the bounds on the emulator training set.}
\label{tab:para}
\end{table}

Galaxy mock catalogs are constructed from the \textsc{Dark Quest} halo catalogs using the halo occupation distribution (HOD) framework, which is implemented in {\tt Halotools} \citep{halotools}. We use the functional form of HOD as proposed in \cite{Zheng07} to model the mean number $\langle N(M)\rangle$ of galaxies in halos of mass $M$. The mean occupation functions of central and satellite galaxies are parameterized as 
\ba \label{eq:hod}
\langle N_{\rm c}(M) \rangle &=& \frac{1}{2} \left[ 1 + \text{erf} \left(
\frac{\log M - \log M_{\text{min}}}{\sigma_{\log M}} \right) \right], \\
\langle N_{\rm s}(M) \rangle &=&  \langle N_{\rm c}(M) \rangle  \left(\frac{M - M_0}{M_1}\right)^{\alpha},
\ea
and $\langle N_{\rm s}(M) \rangle =0$ when $M < M_0$. $M_{\text{min}}$ is the cutoff halo mass scale for hosting central galaxies, with $\langle N_{\rm c}(M_{\rm min})\rangle=0.5$. $\sigma_{\log M}$ describes the profile for the halo mass cutoff, making $\langle N_{\rm c}(M)\rangle$ smoothly transit from 0 to 1. $M_0$ is the minimum halo mass to host satellite galaxies. $M_1$ is the normalization mass scale. $\alpha$ is the power-law slope of the satellite HOD at the massive end. The occupations of central and satellite galaxies are drawn from Bernoulli and Poisson distributions, respectively. Central galaxies are placed at the halo centers with the same velocities as their host halos, where we have ignored the effect of galaxy velocity bias \citep{Guo2015,Guo2016}. We also assume that the satellite galaxy distribution within the halos follows the Navarro-Frenk-White profile \citep{NFW}. 

We adopt the fiducial values of HOD parameters based on the best-fit values ($\log M_{\rm min} = 13.09$, $\sigma_{\log M} = 0.596$, $\log M_0=13.077$, $\log M_1=14.0$ and $\alpha=1.0127$) obtained by fitting to CMASS (``constant mass") galaxy sample \citep{Manera2013}. The number density $n$ can be derived by performing an integral over the mass function,
\ba \label{eq:n}
n = \int dM \frac{dn}{dM}(M) \langle N(M)\rangle,
\ea
where $dn/dM(M)$ is the halo mass function. We take its fitting formula from \citet{Tinker:2008ff}. The resulting HOD catalog has the number density of $n=5.6\times10^{-4}~h^3\,\rm Mpc^{-3}$. In our work, we choose to fix the number density\footnote{Another way is allowing the number density $n$ to vary, then including the information in $n$ by adding a Gaussian prior for $n$ into the likelihood \citep{Lange2021,Donald-McCann:2021nxc}. Varying $n$ would weaken the constraints on HOD parameters to some extent, depending on the uncertainty on $n$, but has a negligible impact on cosmological parameters \citep{Donald-McCann:2021nxc}.}, then we sample four out of the five HOD parameters of our model. Here we re-parameterize HOD parameters as $\{ \sigma_{\log M}, M_0/M_1, \alpha, M_1/M_{\rm min}\}$ as used in \citet{Wibking:2019zuc}. Their fiducial values and flat prior ranges are presented in Table \ref{tab:para}. We utilize the (randomized) quasi-Monte Carlo method to sample re-parameterized HOD parameters in the prior range. Specifically, we generate 2450 points in 4D, using the Sobol sequence \citep{Sobol1967} utility in the \texttt{scipy.stats.qmc} package \citep{Scipy} . We scramble the Sobol sequence with a random seed searched among integers from 0 to 65535 to minimize the mixture discrepancy \citep{Zhou2013MD} as the uniformity measure. The first 2400 HOD samples are assigned to 80 cosmologies for training, \ie\, each training cosmology is assigned 30 HODs. The remaining 50 HODs are assigned to each testing cosmology, yielding a testing set of 1000 models. For each sampling, we use Eq. \ref{eq:n} to find the value of $\log M_{\rm min}$ that yields the fixed $n$. 
We include shifts due to RSD along the $z$-axis to generate the simulated galaxy samples in the redshift space, \ie\, the simulated galaxies in the real space are shifted by $v_z/(aH)$ where $v_z$ is the peculiar velocity of simulated galaxies along line of sight, and $a$ is the scale factor.

\section{Emulating pre- and post-reconstructed galaxy power spectra} \label{sec:emu}
In this section, we use the galaxy samples described in previous section to emulate $P_{\rm pre}, P_{\rm post}$ and $P_{\rm cross}$ of galaxies. We first present the measurement of power spectra with and without the density field reconstruction, then detail the training process of our emulator, and finally discuss the performance of the emulator.

\subsection{The density field reconstruction and the power spectrum measurement}
Before performing the density-field reconstruction, we implement the Alcock-Paczynski (AP) effect \citep{AP}, which arises from the discrepancy between the fiducial cosmology used for redshift-distance conversion and the underlying true cosmology. Although the equation relating the power spectrum before and after applying the AP effect is analytically known \citep{Ballinger:1996cd}, including this effect in the reconstruction is complicated and requires nontrivial modelling \citep{Sherwin:2018wbu}. An easier way to account for the AP effect is to manipulate the catalog by changing the coordinates of the samples. Specifically, we convert the galaxy positions in the true coordinate $\mathbf{x}'$ to the ``observed'' coordinate $\mathbf{x}$, and stretch the size lengths of simulation box $\mathbf{L}$ using the relations of $\mathbf{x} = \mathbf{A}^{-1}\mathbf{x}'$ and $\mathbf{L}\to\mathbf{A}^{-1}\mathbf{L}$ with
\ba
\mathbf{A} = \left(
\begin{array}{ccc}
\alpha_{\perp} & 0 & 0 \\
0 & \alpha_{\perp} & 0 \\
0 & 0 & \alpha_{||} 
\end{array} \right),
\alpha_{\perp} \equiv \frac{D_A(z)}{D_{A, \rm fid}(z)},\alpha_{||} \equiv \frac{H_{\rm fid}(z)}{H(z)}\,,
\ea
where $D_A$ and $H$ are the comoving angular diameter distance and Hubble parameter, and quantities with subscript ``$\rm fid$'' denote those in the fiducial cosmology. The galaxy density field is smoothed by convolving with the kernel $K(k)={\rm exp}\left[- (k\ \Sigma_s)^2/2\right]$ in Fourier space, where $k$ is the modulus of the conjugate wavenumber $\mathbf{k}$ of the observed coordinate $\mathbf{x}$, and we set the smoothing scale to be $\Sigma_s=10~h^{-1}~{\rm Mpc}$, which is close to the optimal smoothing scale for the reconstruction efficiency  \citep{Seo:2015eyw,Vargas-Magana:2015rqa}. The displacement field is then estimated using the Zeldovich approximation, \ie, $\tilde{\mathbf{s}}(\mathbf{k})=-i\frac{\mathbf{k}}{k^2}\frac{\delta({\mathbf{k}})}{b+f \mu^2}K(k)$, where $\delta({\mathbf{k}})$ denotes the nonlinear redshift-space galaxy overdensity in the observed coordinate, $b_{\rm in}$ is the input linear bias of the galaxy sample, and $f_{\rm in}$ is the input logarithmic growth rate. An inverse Fourier transformation on $\tilde{\mathbf{s}}$ returns the configuration-space shift field $\mathbf{s}(\mathbf{x})$, which is used to move both galaxies and randoms. Although it is natural to use the true (fiducial) values of $b$ and $f$ as $b_{\rm in}$ and $f_{\rm in}$ for the reconstruction, this does not have to be the choice. Actually, the true values of $b$ and $f$ are not known before performing the analysis. As we shall demonstrate later, the final parameter estimation is largely insensitive to the choice of $b_{\rm in}$ and $f_{\rm in}$. In what follows, we use the fiducial $b$ and $f$ to start with, and repeat the analysis with a significantly different set of $b_{\rm in}$ and $f_{\rm in}$ to demonstrate the robustness of the final result against the choice of these input parameters.

We measure the multipoles of $P_{\rm pre}, P_{\rm post}$ and $P_{\rm cross}$ using a fast Fourier transform (FFT)-based estimator \citep{Hand17} implemented in {\tt nbodykit} \citep{nbodykit}. The number density field of galaxies is constructed using the cloud-in-cell (CIC) scheme to assign galaxies to the grid, and we correct for the aliasing effect using the interlacing scheme in \citet{Sefusatti:2015aex}. For the monopole of the auto power spectrum before and after density field reconstruction, the shot-noise is removed as a constant. Note that the shot-noise of $P_{\rm cross}$ is scale-dependent \citep{Wang:2022nlx}, which is estimated using the ``half-sum (HS) half-difference (HD)" approach and then subtracted off, as in \citet{noise, Wang:2022nlx}. The $k$-bin width is set to be $\Delta k = 0.01~h~{\rm Mpc}^{-1}$ for all $P(k)$ measurements.

\subsection{Emulating the power spectra}
In order to avoid the emulated quantities spanning several orders of magnitude, we choose to normalize the power spectrum multipoles using the linear Kaiser power spectrum \citep{Kaiser1987} with the BAO feature removed in the fiducial cosmology, \ie\,
\ba \label{eq:R}
R^{\rm X}_0&=& \frac{P_0^{\rm X}}{\left(b^2+2/3bf+1/5f^2 \right)P_{\rm nw,lin}}\,,\\
R^{\rm X}_2&=& \frac{P_2^{\rm X}}{\left(4/3bf+4/7f^2 \right)P_{\rm nw,lin}}\,,\\
R^{\rm X}_4&=& \frac{P_4^{\rm X}}{\left(8/15f^2 \right)P_{\rm nw,lin}}\,,
\ea
where $P_{\rm nw,lin}$ is the linear power spectrum without the BAO feature \citep{Eisenstein:1997ik}. The superscript ${\rm ``X"}$ runs for ``\{pre, post, cross\}". To well capture the BAO wiggles in the monopole, we decompose $R^{\rm X}_0$ into two parts, \ie\, the smoothed broadband shape ($S$) part and the BAO wiggles ($W$) part. The $S$ part is obtained by applying a Savitzky-Golay filter \citep{SGfilter} to $R^{\rm X}_0$, \ie\, fitting to a certain number of data points ($N$) with a polynomial of $p$-th order, and we find that $N=41$ and $p=4$ is a reasonable choice for the filtering. Then the BAO wiggles are extracted, \ie\, $W^{\rm X}_0 = R^{\rm X}_0-S^{\rm X}_0$. Fig. \ref{fig:trainR} in the Appendix shows the observables ($2400$ in total) used for training the emulator.

\begin{figure*}[!t]
\centering
\includegraphics[scale=0.27]{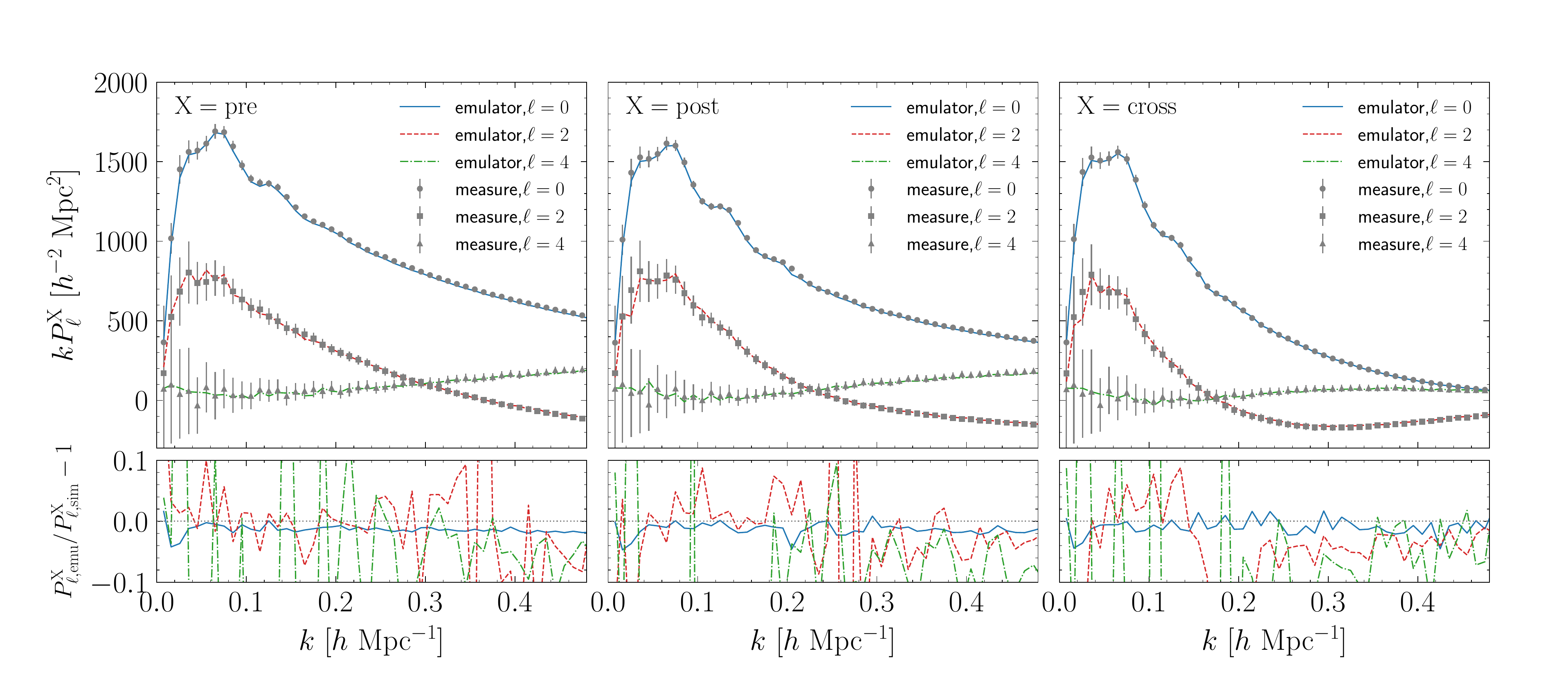}
\caption{Upper panels: The prediction from our emulator for multipole moments of $P_{\rm pre}, P_{\rm post}$ and $P_{\rm cross}$ for the fiducial cosmology that is not used for the training. The symbols are the average of $15$ realizations in the fiducial cosmology. The errors are the statistical errors for a volume of $3~h^{-3}~\rm Gpc^3$. Lower panels: The fractional difference between the emulator prediction and the measured power spectra from mocks in the fiducial cosmology.}
\label{fig:fidpred}
\end{figure*} 

We follow \citet{Zhai:2018plk, Zhai_2023} to construct the emulator, based on the {\tt George} (\citealt{george} code. In the GP modelling, the correlation between different training data points is modelled by a covariance matrix generated by a kernel function. This is of critical importance in the GP modelling since it defines the function we wish to learn. Due to the lack of prior knowledge of the correlation between training data points, the definition of the kernel function can be arbitrary. For the modelling of galaxy power spectrum in this work, we adopt a Matern class kernel ($\mathcal{K}$) as we find it produces sufficiently accurate predictions. In this model, the hyperparameters in the kernel define the strength of correlation between neighboring points. The following process of training is to optimize the hyperparameters in the kernel function:
\ba
\ln \mathcal{L}=-\frac{1}{2} \bold P^T M^{-1} \bold P-\frac{1}{2} \ln |M|-\frac{1}{2} N \ln 2 \pi, \,
\ea
where $M=\mathcal{K} + \sigma^2 I$, $\mathcal{K}$ is the covariance matrix populated by the kernel function, and $\sigma$ represents the error of the training data $\bold P$. Since each cosmology in the training data has only one realization, we estimate the uncertainty of the training data using the fiducial cosmology with $15$ realizations. We find $15$ simulations gives a good approximation of $\sigma$, but more simulations might improve the emulator, which would be interesting to check in the future. With the optimized hyperparameters fed into the kernel function, we can obtain the power spectra for an arbitrary point in the parameter space.

\begin{figure*}[!t]
\centering
\includegraphics[scale=0.45]{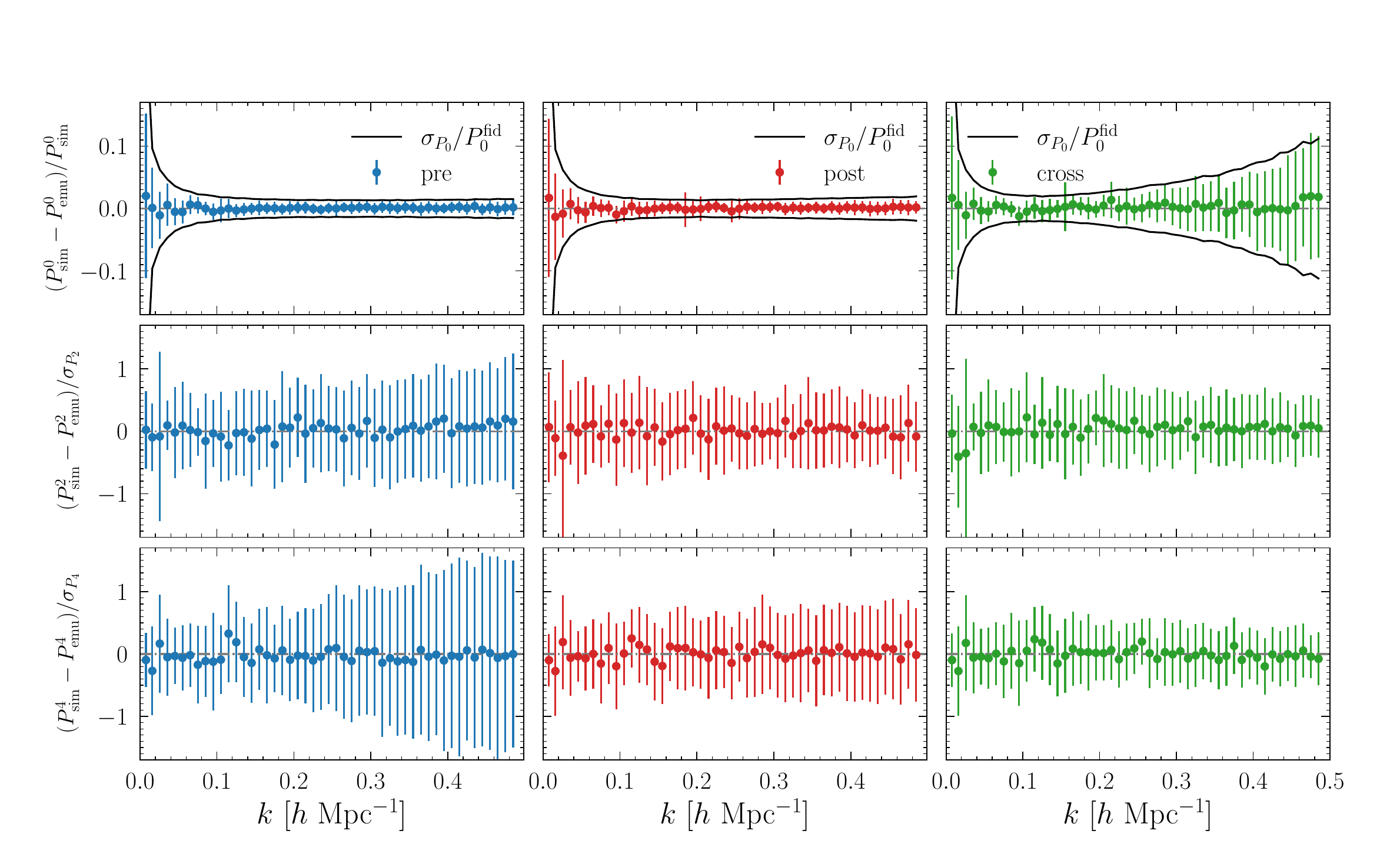}
\caption{Upper panels: The average of the fractional difference between the emulated and the measured monopole power spectrum from $1000$ testing mocks. The black solid lines show the inverse signal-to-noise ratio of the mean fiducial monopole measurement. The statistical error for monopole power spectrum $\sigma_{P_0}$ is computed using Eq.\,\ref{eq:Cdata}. Middle panels: The average of the difference between the emulated and the measured quadrupole power spectrum relative to the statistical error. Lower panels: The average of the difference between the emulated and the measured hexadecapole power spectrum relative to the statistical error.}
\label{fig:emuaccuracy}
\end{figure*}

\subsection{Covariance matrix} 
\label{subsec:Cdata}
The \textsc{Dark Quest} simulations only have $15$ realizations in the fiducial cosmology, which is insufficient to construct a robust covariance matrix for galaxy clustering analysis. We therefore compute the correlation matrix using GLAM simulations, for which there are $986$ independent realizations in the Planck cosmology\footnote{The cosmological parameters of GLAM simulations are from Planck 2013 \citep{Planck:2013pxb}, which is slightly different from the fiducial cosmology used in the \textsc{Dark Quest} simulations. The minor difference between them is ignored in this work.} \citep{GLAM}. We adopt the best-fit HOD parameters for the $M_i < -21.6$ CMASS samples in \citet{Guo:2014oca}, leading to a contribution of shot noise ($\sim 2 \times 10^{-4}~h^3~\rm Mpc^{-3}$) to the covariance. The side length of the GLAM simulation box is $1~h^{-1}~\rm Gpc$. Simulations with larger box size can help investigate the effect of super-sample covariance (SSC) \citep{Bayer:2022nws}. Limited by the simulation suite, the covariance determination hence in this work neglects super-sample variance components\footnote{\citet{Hikage2020} found the improvement of the error on the growth rate by the reconstruction was comparable between the cases with and without the SSC effect. It would be interesting to explore the impact of the SSC on cosmological parameters in future work.}. To be close to the volume of the BOSS survey \citep{Alam:2016hwk}, the data covariance matrix $C_{\rm data}$ is rescaled by a factor of $3$. Specifically, we derive the $C_{\rm data}$ from GLAM mocks, \ie
\ba \label{eq:Cdata}
(C^{\ell, \ell'}_{i,j})_{\rm data}&=& \frac{1}{N_s-1}\sum_{n=1}^{N_s} \left[P_{\ell}^{n}(k_i)-\overline{P_{\ell}}(k_i)\right] \times \nonumber \\
&&\left[P_{\ell'}^{n}(k_j)-\overline{P_{\ell'}}(k_j)\right] \,,
\ea
where the mean of power spectra is defined as
\ba
\overline{P_{\ell}}(k_i) = \frac{1}{N_s}\sum_{n=1}^{N_s} P_{\ell}^{n}(k_i)\,,
\ea
and $N_s=986$ is the number of mocks. Note that the effect of the error induced by the estimation of covariance matrix from the finite number of mocks will be corrected as described later in this work.

\subsection{Emulator validation} 
\label{subsec:validation}

In Fig. \ref{fig:fidpred}, we show the prediction from our emulator for multipole moments of $P_{\rm pre}, P_{\rm post}$ and $P_{\rm cross}$ for the fiducial cosmology that is not used for the training. The symbols in the upper panels are the average of power spectra measured from $15$ realizations in the fiducial cosmology. The error bars are the statistical errors computed using Eq.\,\ref{eq:Cdata}. 

The lower panels of Fig. \ref{fig:fidpred} show the fractional difference between the emulated and the measured power spectra from the galaxy mocks. It indicates that the monopole and quadrupole measured from the galaxy mocks can be well described by our emulator, by better than 1-2$\%$ for the monopole and 2-10$\%$ for the quadrupole at most scales. The quadrupole error can sometimes be $>10\%$, particularly around the scales where the quadrupole happens to cross zero.  For the hexadecapole, the fractional difference is noisy because the amplitudes of hexadecapole are close to zero. Within the statistical errors, our emulator gives an excellent prediction for the hexadecapole as well. 

We quantify the accuracy of our emulator using $1000$ test galaxy mocks that are not used in the training set. The three columns of Fig. \ref{fig:emuaccuracy} from left to right show the performance of our emulator for $P_{\rm pre}$ (left), $P_{\rm post}$ (middle) and $P_{\rm cross}$ (right), respectively.

The symbols in the upper panels of Fig. \ref{fig:emuaccuracy} show the average fractional error of the monopole power spectrum obtained by comparing the emulator predictions with the $P_0$ measurements from $1000$ test mocks. The error bars are the standard deviation estimated from $1000$ test mocks. The fractional error is within $\sim 1$-$2\%$ over most scales. The solid lines show the inverse signal-to-noise ratio computed using the average of the monopole measurements among $15$ realizations from the fiducial cosmology. 

Because the quadrupole and hexadecapole moments can cross zero, leading to large fractional errors, we instead show the absolute error between the emulator prediction and the measurement from the testing mocks, relative to the statistical error in the middle and lower panels of Fig. \ref{fig:emuaccuracy}. We find that the emulator error for $P_2$ and $P_4$ is sub-dominant, roughly $50-70\%$ of the statistical error for a volume of $3~h^{-3}~\rm Gpc^3$.

\begin{figure*}[!t]
\centering
\includegraphics[scale=0.55]{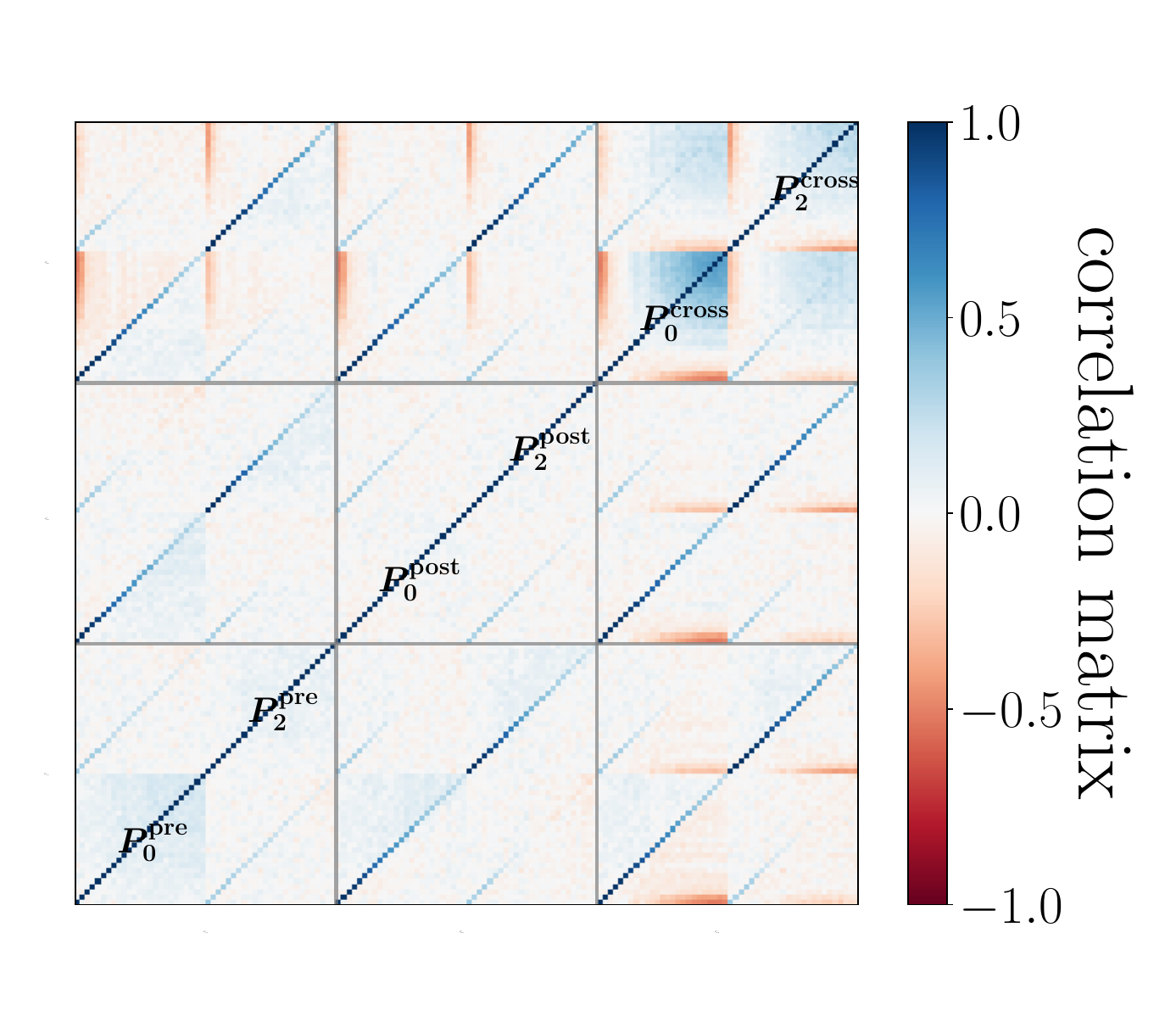}
\caption{The correlation matrix for the monopoles and quadrupoles of power spectra $(P_{\rm pre}, P_{\rm post}, P_{\rm cross})$.}
\label{fig:corrmat}
\end{figure*}

\begin{table*}[!t]
\centering
\footnotesize
\begin{tabular}{c|c|cccc|c|c}
\hline\hline
             & \multicolumn{6}{c|}{$k_{\rm max}=0.25$} &$k_{\rm max}=0.5$
             \\\hline
$b_{\rm in}, f_{\rm in}$            && \multicolumn{4}{c|}{$(b_{\rm fid}, f_{\rm fid})$} & \multicolumn{1}{c|}{$(0.9\,b_{\rm fid}, 0.7\,f_{\rm fid})$}& $(b_{\rm fid}, f_{\rm fid})$\\\hline
             &  $P_{\rm pre}$ & $ P_{\rm post}$ & $P_{\rm cross}$  & $P_{\rm pre+post}$ & $P_{\rm all}$ & $P_{\rm all}$ & $P_{\rm post}$ 
             \\\hline
          $ \mathcal{P}$-$\rm factor$    &  $1.09$ & $1.09$ & $1.09$  & $1.22$ & $1.36$ & $1.36$ & $1.21$
            \\\hline
$\Omega_m$  & $0.318 \pm 0.0110 $ & $0.315 \pm 0.0098$ & $0.317 \pm 0.0100$ & $0.318 \pm 0.0082 $& $0.320 \pm 0.0061$ &$0.320 \pm 0.0080$ &$ 0.317 \pm 0.0078 $ \\ 
$H_0$  & $67.13 \pm 0.84 $ &$67.00\pm0.54$ &$67.05 \pm 0.69$ & $67.15 \pm 0.52$ & $ 67.06 \pm 0.49$ &$67.12 \pm 0.49 $&$ 67.02 \pm 0.57$  \\ 
$\sigma_8$  &$0.849 \pm 0.038$ &$0.833 \pm 0.029$ & $0.842 \pm 0.040$ & $0.834 \pm 0.018$ & $0.834 \pm 0.017$ &$ 0.834 \pm 0.018$&$0.831 \pm 0.016 $\\ \hline
$\alpha_{\perp}$  & $1.0008\pm0.0115$ & $1.0039 \pm 0.0088$& $1.0022 \pm 0.0097$&  $1.0004 \pm 0.0083 $ & $1.0013 \pm 0.0076$ &$1.0003 \pm 0.0081 $&$1.0028 \pm 0.0090 $\\
$\alpha_{||}$  & $1.0000\pm0.0118 $ & $1.0041 \pm 0.0106 $& $1.0016 \pm 0.0104$& $0.9996 \pm 0.0097$& $0.9999 \pm 0.0084$ &$0.9987 \pm 0.0098$&$ 1.0024 \pm 0.0103$\\
$f\sigma_8$  & $0.497 \pm 0.023$ &  $0.487 \pm 0.018$ &$ 0.492 \pm 0.024$ &  $0.488 \pm 0.0116$ & $0.488 \pm 0.0104 $ &$0.488 \pm 0.0114 $&$0.486 \pm 0.010$ \\
\hline\hline
\end{tabular}
\caption{The constraints on derived cosmological parameters $(\Omega_m, H_0, \sigma_8)$ and BAO and RSD parameters $(\alpha_{\perp}, \alpha_{||}, f\sigma_8)$ using different data sets. The fiducial values of the parameters derived are $\Omega_m=0.3156$, $H_0=67.24$, $\sigma_8=0.831$, $\alpha_{\perp}=1$, $\alpha_{||}=1$ and $f\sigma_8(z=0.549)=0.485$, which are well recovered in all cases. The factor $\mathcal{P}$ here is calculated using Eq.\ref{eq:covfactor}. Our default choices of $(b, f)$ parameters for reconstruction are $b_{\rm fid} =1.824$ and $f_{\rm fid} =0.778$ determined in the fiducial cosmology. To explore the effect of these inputs, we vary the bias by $-10\%$ (\ie\,$0.9\,b_{\rm fid}$) and the $f$ by $-30\%$ (\ie\, $0.7\,f_{\rm fid}$).}
\label{tab:result}
\end{table*}

\begin{figure*}[!t]
\centering
\includegraphics[scale=0.55]{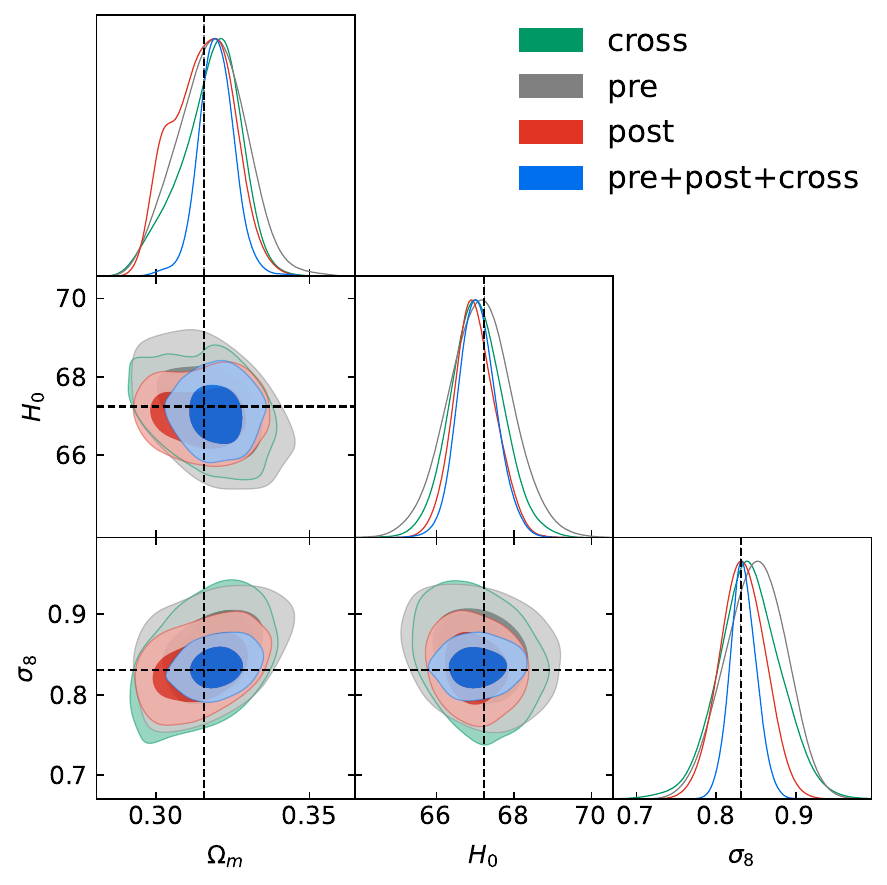}
\includegraphics[scale=0.55]{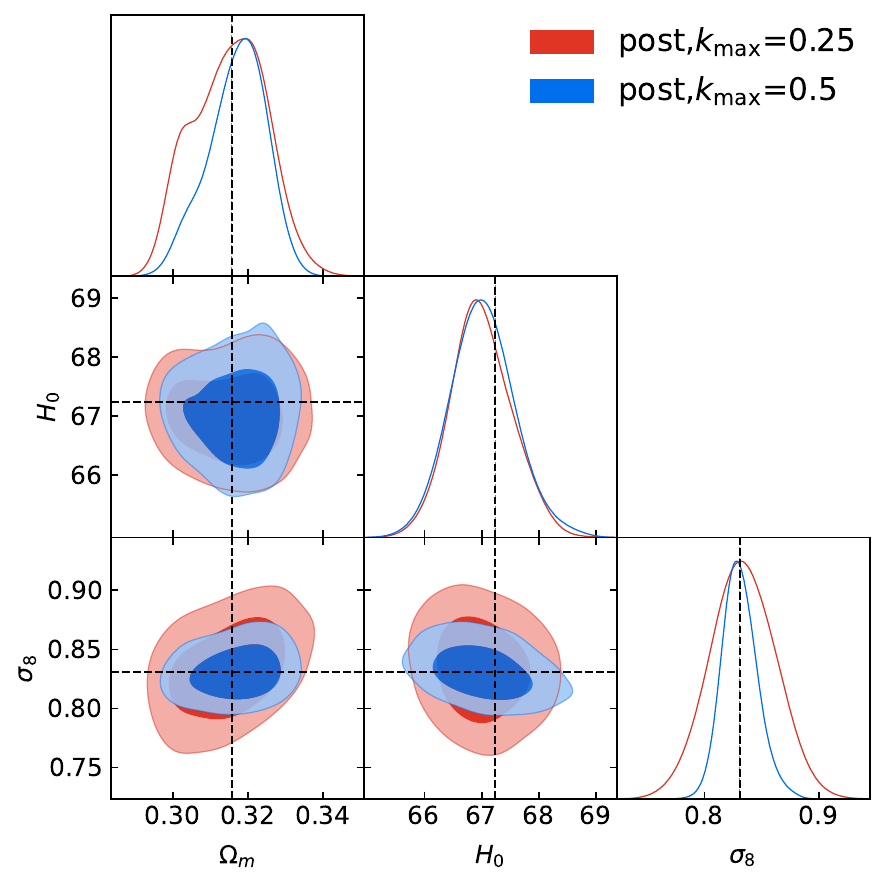}
\caption{Left: The 1D posterior distribution and 2D contour plots showing 68\% and 95\% credible regions for $\Omega_m$, $H_0$ and $\sigma_8$ using the pre-reconstructed power spectrum alone (grey), post-reconstructed power spectrum alone (red), cross power spectrum alone (green), and joint result of pre, post and cross power spectra (blue). Right: The same plot derived from the post-reconstructed power spectrum alone with two choices of $k_{\rm max} =0.25~h~\rm Mpc^{-1}$ (red) and $k_{\rm max} =0.5~h~\rm Mpc^{-1}$ (blue). 
}
\label{fig:cos}
\end{figure*}

\begin{figure*}[!t]
\centering
\includegraphics[scale=0.55]{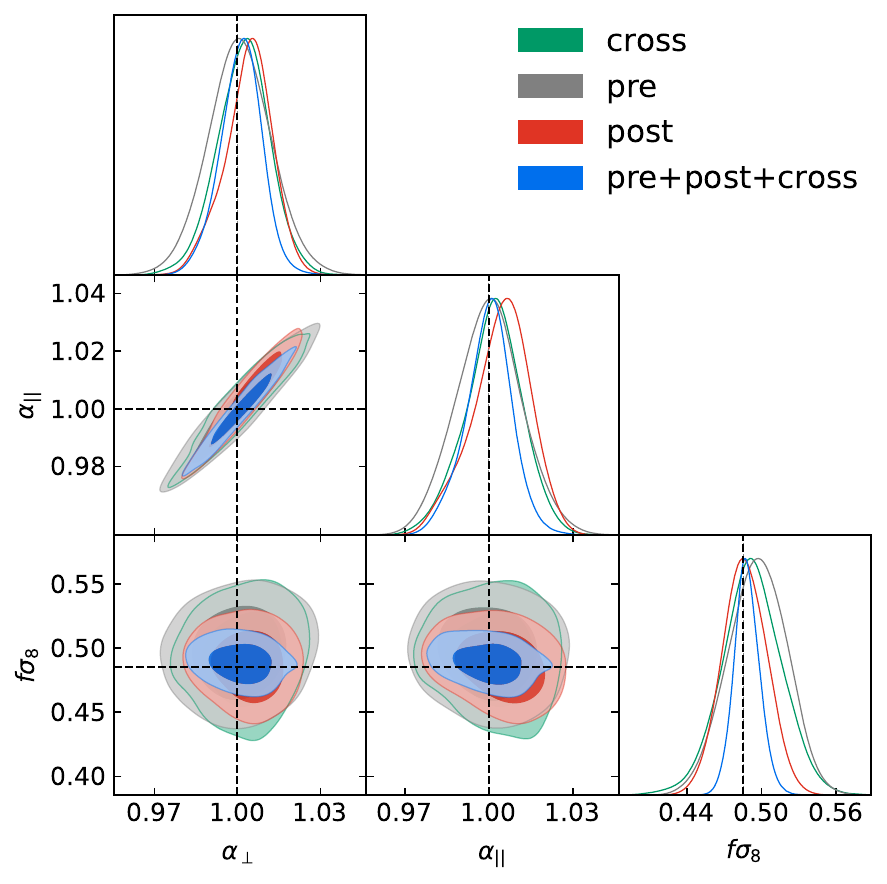}
\includegraphics[scale=0.55]{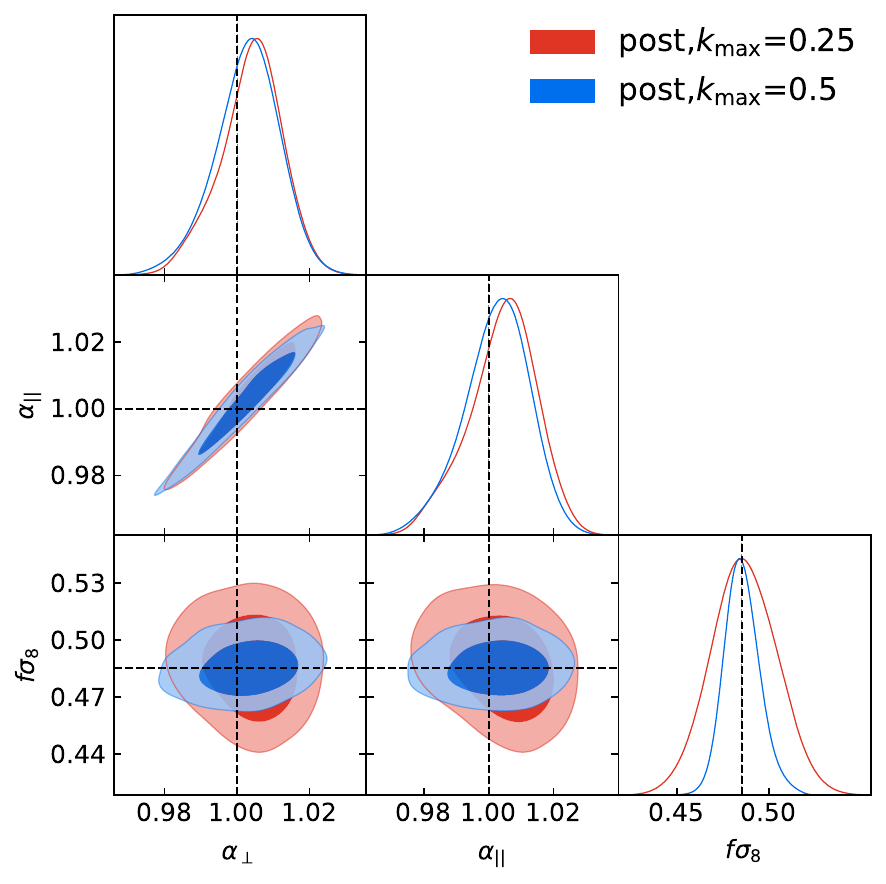}
\caption{Left: The 1D posterior distribution and 2D contour plots showing 68\% and 95\% credible regions for the derived $\alpha_{\perp}$, $\alpha_{||}$ and $f\sigma_8$ parameters using the pre-reconstructed power spectrum alone (grey), post-reconstructed power spectrum alone (red), cross power spectrum alone (green), and joint result of pre-, post- and cross-power spectra (blue). Right: The result using the post-reconstructed power spectrum alone with two choices of $k_{\rm max} =0.25~h~\rm Mpc^{-1}$ (red) and $k_{\rm max} =0.5~h~\rm Mpc^{-1}$ (blue).}
\label{fig:BAORSD}
\end{figure*}

\section{Cosmological application to mock catalogs} \label{sec:fit}
In this section, we test our emulator by applying it to the power spectrum measurements from mock galaxy catalogs in the fiducial cosmology, which are not in the training set. We use {\tt Cobaya} \citep{Torrado:2020dgo} to perform a Markov chain Monte Carlo (MCMC) sampling of the 9-dimensional parameter space within the flat $\Lambda$CDM framework, \ie\, the $w$ parameter is fixed to $-1$. The following $\chi^2$ gets minimised in the fitting,
\ba
 \chi^2 \equiv \left[P_{\rm emu} (k) -P_{\rm mea}(k) \right]^T C^{-1} \left[P_{\rm emu} (k) -P_{\rm mea}(k) \right]\,,
\ea
and we add a Gaussian prior for $\omega_b$ centered on $0.0223$ with the width $0.00036$ from BBN constraints \citep{Mossa:2020gjc} and a Gaussian prior for $n_s$ parameters centered on $0.965$ with the width $0.0042$ from Planck constraints \citep{Planck:2018vyg}. For other parameters, a flat prior over the range shown in the Table\,\ref{tab:para} is used. Here $P_{\rm emu}$ is the prediction of the emulator, and $P_{\rm mea}$ denotes the average of power spectra measured from $15$ realizations in the fiducial cosmology. $C$ is the covariance matrix consisting of two terms,
\ba
C = C_{\rm data} + \sigma^2_{\rm emu}I\,,
\ea 
where $C_{\rm data}$ (computed in Sec.\,\ref{subsec:Cdata}) is the contribution of the sample statistics, and $\sigma_{\rm emu}$ corresponds to the uncertainty due to the emulating error in the model prediction. Since the emulator is constructed for individual scale bins, here we assume that emulating error is independent among different scale bins, which is computed using the testing set as discussed in Sec.\,\ref{subsec:validation}. Since the covariance matrix $C_{\rm data}$ is estimated from finite number of mocks, $C_{\rm data}$ is generally biased. To correct, we multiply $C_{\rm data}$ by a factor of $\mathcal{P}$ \citep{Percival:2021cuq}, 
\ba \label{eq:covfactor}
\mathcal{P} = \frac{(N_s-1)[1+B(N_d-N_{\theta})]}{N_s-N_d+N_{\theta}-1}\,,
\ea
with,
\ba
B=\frac{N_s-N_d-2}{(N_s-N_d-1)(N_s-N_d-4)}\,. 
\ea
Here, $N_s$ is the number of simulations used to estimate the covariance, $N_d$ is the number of the data vector, and $N_{\theta}$ is the number of parameters that are being fitted. Note that the $\mathcal{P}$-factor generally dilutes the constraints on parameters being fitting by rescaling the covariance, namely, when using $P_{\rm pre}$, $P_{\rm post}$, or $P_{\rm cross}$ alone, the $\mathcal{P}$-factor increases the covariance by $9\%$. For joint analyses of $P_{\rm pre}+P_{\rm post}$ and $P_{\rm pre}+P_{\rm post}+P_{\rm cross}$  (\ie\,$P_{\rm all}$), the $\mathcal{P}$-factor enlarges the covariance by $22\%$ and $36\%$, respectively. These $\mathcal{O}$(10\%) enlargements of the posteriors are a first-order way to correct for the lack of convergence of the covariance due to using a small number of simulations. Running more simulations would reduce the size of this correction and provide more robust contours. We plan to do this in future work. When a covariance matrix is constructed using Eq.~\ref{eq:Cdata} we have, in effect, drawn the matrix as a random variable from a Wishart distribution. Using this knowledge, we can consider the set of covariance matrices that could have been drawn, and determine the average effects on results from them. It is often found that the results are biased because the values of interest are skewed by the errors in the covariance. Examples of such effects include a skewed inverse covariance matrix \citep{Hartlap:2006kj} or skewed parameter errors \citep{Dodelson:2013uaa}. The effects can be modelled using perturbation theory (PT), leading to correction terms such as those in Eq.~\ref{eq:covfactor}. The PT-based derivation assumes a linear model, but this has been shown to work well for typical cosmological problems \citep{Percival:2021cuq}. Nevertheless, this is only a first-order correction in terms of the link from the likelihood to model parameters, and having more mocks will always be better. The correlation matrix being estimated for the combined power spectra ($P_{\rm pre}, P_{\rm post}, P_{\rm cross}$) is shown in Fig.~\ref{fig:corrmat}.

Using $k$ modes at $k\le0.25~h~\rm Mpc^{-1}$ for both the monopole and quadrupole of the power spectra\footnote{Unless otherwise mentioned, $k_{\rm max}$ is $0.25~h~\rm Mpc^{-1}$ as a default setting for the analysis in this work.}, we obtain the 1D posterior distributions and 2D contour plots for the derived cosmological parameters $\Omega_m$, $H_0$ and $\sigma_8$, as shown in Fig. \ref{fig:cos}. The mean values with 68\% credible intervals of $\Omega_m$, $H_0$ and $\sigma_8$ are presented in Table \ref{tab:result}. The left contour plot in Fig. \ref{fig:cos} shows a comparison of the fitting results using the pre-reconstructed power spectrum alone (grey), post-reconstructed power spectrum alone (red), cross power spectrum alone (green), and the joint fitting of pre-, post-, and cross-power spectra (blue) for $k_{\rm max} =0.25~h~\rm Mpc^{-1}$. Fig. \ref{fig:cos} shows that our emulator-based analysis can recover the expected values of cosmological parameters within statistical errors. The post-reconstructed power spectrum alone is more informative, tightening the constraints on $\Omega_m$, $H_0$ and $\sigma_8$ by $10.9\%$, $35.7\%$ and $23.7\%$, respectively compared to that from $P_{\rm pre}$ alone. It is found that the joint fit of the pre-, post-, and cross-power spectra, denotes as $P_{\rm all}$, gives the tightest constraint, namely, the constraints on $\Omega_m$, $H_0$ and $\sigma_8$ from $P_{\rm all}$ are improved by $44.5\%$, $41.7\%$, and $55.3\%$, respectively, compared to that from $P_{\rm pre}$ alone. 

The relative information gain from $P_{\rm all}$ compared to that from $P_{\rm pre}$ is expected to be greater in the nonlinear regime, \ie\, including modes with $k>0.25~h~\rm Mpc^{-1}$. However, given the number of mocks and data points, we do not go further than $k_{\rm max}= 0.25~h~\rm Mpc^{-1}$ for a $P_{\rm all}$ analysis. Instead, we perform a $P_{\rm post}$-alone analysis with $k_{\rm max}= 0.5~h~\rm Mpc^{-1}$ for a demonstration. We compare the constraints on $\Omega_m$, $H_0$ and $\sigma_8$ using $P_{\rm post}$ alone for $k_{\rm max}= 0.25$ and $0.5~h~\rm Mpc^{-1}$ in the right panel of Fig. \ref{fig:cos}. As shown, adding modes on smaller scales helps to constrain $\sigma_8$, namely, its uncertainty gets reduced by $44.8\%$ as $k_{\rm max}$ increases from $0.25$ to $0.5~h~\rm Mpc^{-1}$. The galaxy clustering on smaller scales is more sensitive to the amplitude-related parameter $\sigma_8$, compared to $\Omega_m$ and $H_0$. In addition, we perform the analysis for different $k_{\rm max}$ using $P_{\rm post}$ alone. The fitting results as a function of $k_{\rm max}$ are presented in Fig. \ref{fig:kmax} in the Appendix. Also, adding more modes does not generate bias in the posteriors, demonstrating the robustness of our emulator.

We then derive the BAO and RSD parameters ($\alpha_{\perp}$, $\alpha_{||}$, $f\sigma_8$), and show the 1D posterior distributions and 2D contour plots in Fig. \ref{fig:BAORSD}, with the mean values and 68\% credible intervals of the BAO and RSD parameters listed in Table \ref{tab:result}. Compared to $P_{\rm pre}$ alone, the constraints on ($\alpha_{\perp}$, $\alpha_{||}$, $f\sigma_8$) parameters from $P_{\rm all}$ are improved by $33.9\%$, $28.8\%$, and $54.8\%$, respectively. $P_{\rm post}$ alone gives a tighter constraint than that using $P_{\rm pre}$ only, but is outnumbered by $P_{\rm all}$ by $13.6\%$ for $\alpha_{\perp}$, $20.8\%$ for $\alpha_{||}$ and $42.2\%$ for $f\sigma_8$. The right panel of Fig. \ref{fig:BAORSD} shows the contours of the derived BAO and RSD parameters with two different choices of $k_{\rm max}$, as in Fig. \ref{fig:cos}. As expected, adding small-scale modes ($k\in[0.25,0.5]~h~\rm Mpc^{-1}$) helps to tighten the constraint on $f\sigma_8$ significantly, namely, the uncertainty gets reduced by $44.4\%$. Note that this level of constraint can be achieved by using $P_{\rm all}$ with $k_{\rm max}=0.25~h~\rm Mpc^{-1}$.

We confirm that the information content in $P_{\rm cross}$ is complementary to that in $P_{\rm pre}$ and $P_{\rm post}$, as claimed in \cite{Wang:2022nlx}. Specifically, adding $P_{\rm cross}$ to our joint analysis using $P_{\rm pre}$ and $P_{\rm post}$ improves the constraints on ($\Omega_m$, $H_0$, $\sigma_8$) and ($\alpha_{\perp}$, $\alpha_{||}$, $f\sigma_8$) by $5.5\%$-$25.6\%$, as presented in Table \ref{tab:result}.

Since the BAO reconstruction process requires a pair of input $b$ and $f$, denoted as $b_{\rm in}$ and $f_{\rm in}$, the reconstructed power spectrum depends on $b_{\rm in}$ and $f_{\rm in}$. One natural question is whether and how much the final posterior depends on $b_{\rm in}$ and $f_{\rm in}$. To investigate, we use a set of $b_{\rm in}$ and $f_{\rm in}$ that are significantly different from the fiducial $b$ and $f$, namely, $b_{\rm in}=0.9\,b_{\rm fid}$ and $f_{\rm in}=0.7\,f_{\rm fid}$. Note that this level of deviation from the true values is much greater than that constrained by a typical galaxy survey such as BOSS \citep{BOSS:2016psr}, thus is sufficient to study the impact of using `wrong' cosmological parameters for the reconstruction on the final result \citep{Sherwin:2018wbu}. We repeat our analysis using this set of $b_{\rm in}$ and $f_{\rm in}$, and show the parameter constraint from $P_{\rm all}$ in this case in Table \ref{tab:result} and in Fig. \ref{fig:all_diff_bf} in the Appendix. As shown, the constraint is largely unchanged, demonstrating the robustness of our method against the choice of $b_{\rm in}$ and $f_{\rm in}$.

For completeness, we show the full contour plot for all parameters, including the cosmological and HOD parameters, in Fig. \ref{fig:all} in the Appendix using different combinations of the power spectra. As expected, $P_{\rm all}$ provides the tightest constraint for all parameters, as predicted by the Fisher matrix analysis \citep{Wang:2022nlx}. 

Results presented so far do not include information from $P_4$, the hexadecapole, so it is useful to explore how $P_4$ can help to reduce the uncertainties. We perform an additional analysis using $P_{\rm all}$ including $P_4$ for all types of power spectra with $k_{\rm max}=0.25~h~\rm Mpc^{-1}$, and find that $P_4$ can barely further improve the constraints on cosmological parameters, as shown in Fig. \ref{fig:hexa} in the Appendix, because the hexadecapole has a relatively lower signal-to-noise ratio compared to $P_0$ and $P_2$. A similar conclusion is found when only using $P_{\rm pre}$, and the contour plot is presented in Fig. \ref{fig:prehexa}.

\section{Conclusion and discussions}
\label{sec:conclusion}
In this work, we develop an emulator for galaxy power spectra for catalogs with and without the BAO reconstruction based on the \textsc{Dark Quest} simulations with HOD models to populate galaxies. The theoretical predictions of power spectra derived from our emulator are in excellent agreement with the ground truth (with a deviation less than $10\%$). Our emulator-based likelihood analysis on mock galaxy catalogs demonstrates that input cosmological parameters can be accurately recovered from power spectra up to scales of $k =0.5~h~\rm Mpc^{-1}$.

Our analysis shows that $P_{\rm pre}$, $P_{\rm post}$ and $P_{\rm cross}$ are highly complementary, thus jointly using these power spectra can significantly improve constraints on cosmological parameters, which is consistent with the claim based on a Fisher matrix analysis \citep{Wang:2022nlx}. Specifically, the uncertainty of $(\Omega_m, H_0, \sigma_8)$ derived from $P_{\rm pre}+P_{\rm post}+P_{\rm cross}$ gets tightened by $44.5\%$, $41.7\%$ and $55.3\%$, respectively, compared to that derived from $P_{\rm pre}$ ($k_{\rm max} = 0.25~h~\rm Mpc^{-1}$ in all cases). The derived BAO and RSD parameters, $\alpha_{\perp},\alpha_{||}$ and $f\sigma_8$, are better determined by $33.9\%,28.8\%$ and $54.8\%$, respectively. Adding small-scale modes to the analysis helps to constrain parameters related to the amplitude of power spectra. For example, extending $k_{\rm max}= 0.25$ to $0.5~h~\rm Mpc^{-1}$ for $P_{\rm post}$ reduces the uncertainty on $\sigma_8$ and $f\sigma_8$ by $44.8\%$ and $44.4\%$, respectively. We also find that the posteriors of parameters are largely insensitive to input values of $b$ and $f$, which are required for the BAO-reconstruction process.

The methodology and pipeline developed in this work make it possible to extract high-order information from two-point statistics, which is of significance for cosmological studies. Our method and emulator can be directly applied to existing and forthcoming galaxy surveys including BOSS \citep{Dawson2013}, eBOSS \citep{Dawson2016}, DESI (Dark Energy Spectroscopic Instrument, \citealt{DESI:2016fyo,DESI:2016igz}), PFS (Prime Focus Spectrograph, \citealt{PFS}) and so forth, after the required tuning in the emulation process for the number density, effective redshifts of the galaxy samples \etc, which is technically straightforward.

\newpage

\acknowledgements
We thank Hee-Jong Seo for helpful discussions. YW is supported by National Key R\&D Program of China (2022YFF0503404, 2023YFA1607800, 2023YFA1607803), NSFC Grants 12273048, 11890691, 11720101004, the CAS Project for Young Scientists in Basic Research (No. YSBR-092), and the Youth Innovation Promotion Association CAS. RZ, YF, YW and GBZ are supported by NSFC grants 11925303 and 11890691. RZ is also supported by the Chinese Scholarship Council (CSC) and the University of Portsmouth. ZZ is supported by NSFC(12373003), and acknowledges the generous sponsorship from Yangyang Development Fund. KK is supported by the STFC grant ST/W001225/1. HG is supported by National SKA Program of China (No. 2020SKA0110100), NSFC Grants 11922305, 11833005, and the science research grants from the China Manned Space Project with No. CMS-CSST-2021-A02. GBZ is also supported by the CAS Project for Young Scientists in Basic Research (No. YSBR-092), the science research grants from the China Manned Space Project with No. CMS-CSST-2021-B01, and the New Cornerstone Science Foundation through the XPLORER prize. TN is supported by MEXT/JSPS KAKENHI Grant Numbers JP19H00677, JP20H05861, JP21H01081 and JP22K03634. Numeric work was performed on the UK Sciama High Performance Computing cluster supported by the ICG, University of Portsmouth. YW thanks the ICG, University of Portsmouth for their hospitality. 

\bibliography{ms}{}
\bibliographystyle{aasjournal}

\newpage
\appendix

This appendix includes four figures, with information detailed in the figure captions.

\begin{figure*}[!htp]
\centering
\includegraphics[scale=0.45]{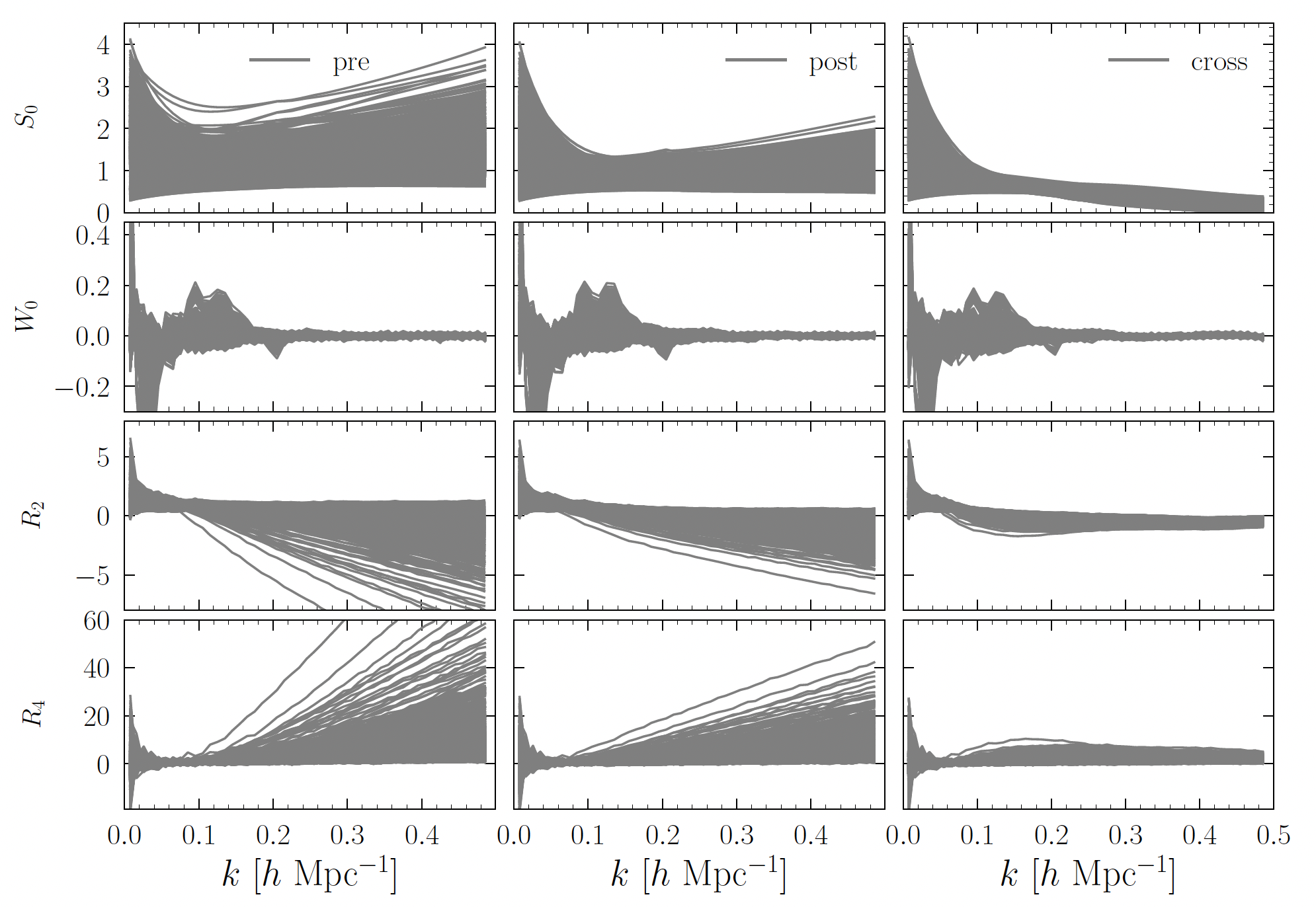}
\caption{The complete training set for our emulator, which consists of $2400$ power spectrum multipoles for $P_{\rm pre}$ (left column), $P_{\rm post}$ (middle) and $P_{\rm cross}$ (right column). All spectra have been properly normalized by power spectra derived from the linear Kaiser formula, so that their amplitudes are within a narrow range. The normalized monopole, $R_0^{\rm X}$, is divided into a smoothed shape part ($S_0^{\rm X}$) and a BAO ``wiggles" part ($W_0^{\rm X}$). More details are presented in the main text and Eq. (\ref{eq:R}).}
\label{fig:trainR}
\end{figure*}

\begin{figure*}[!t]
\centering
\includegraphics[scale=0.55]{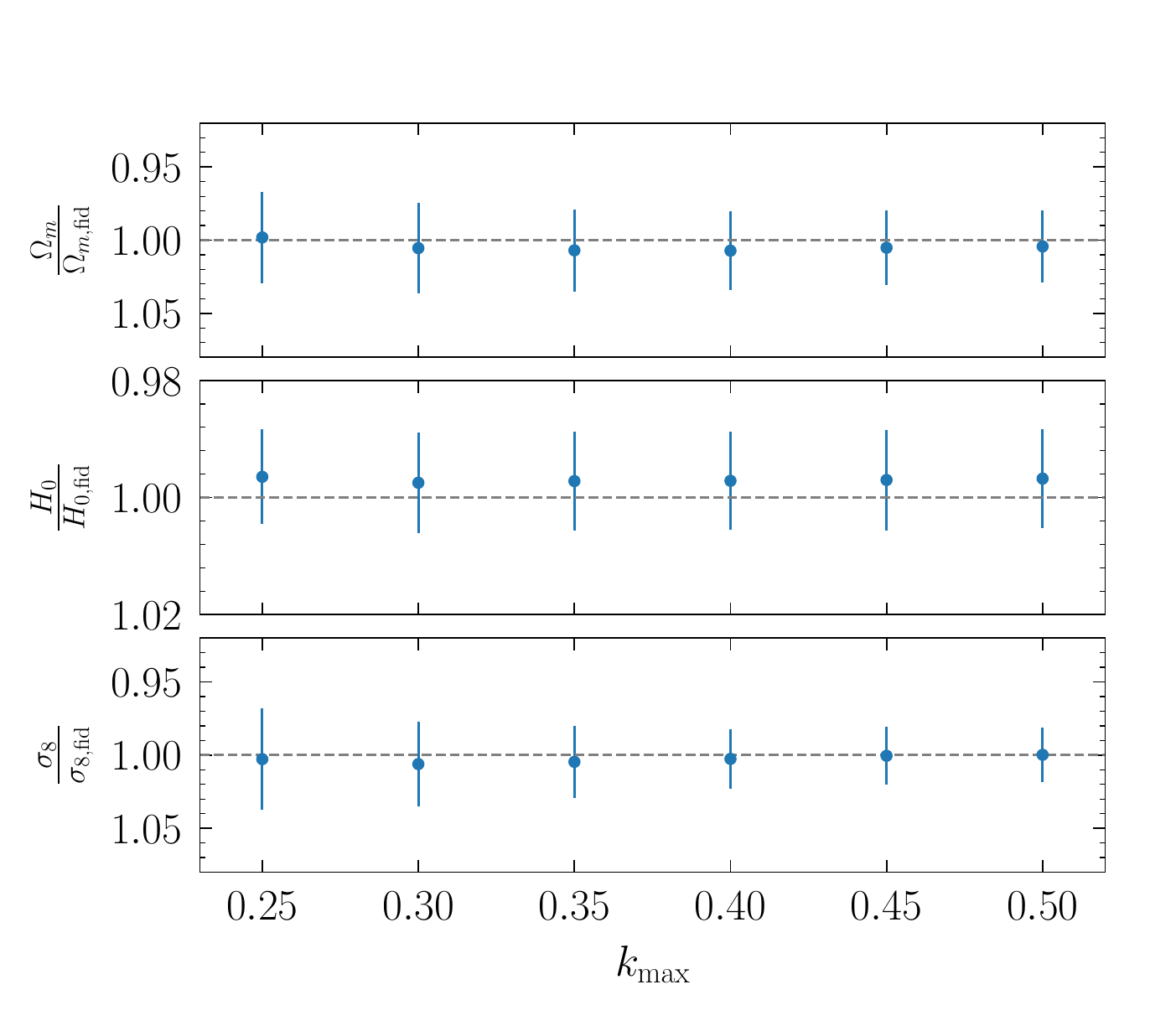}
\caption{The mean and standard deviation of ($\Omega_m$, $H_0$, $\sigma_8$) parameters normalized by the fiducial values as a function of $k_{\rm max}$ using $P_{\rm post}$ alone.}
\label{fig:kmax}
\end{figure*}

\begin{figure*}[!t]
\centering
\includegraphics[scale=0.55]{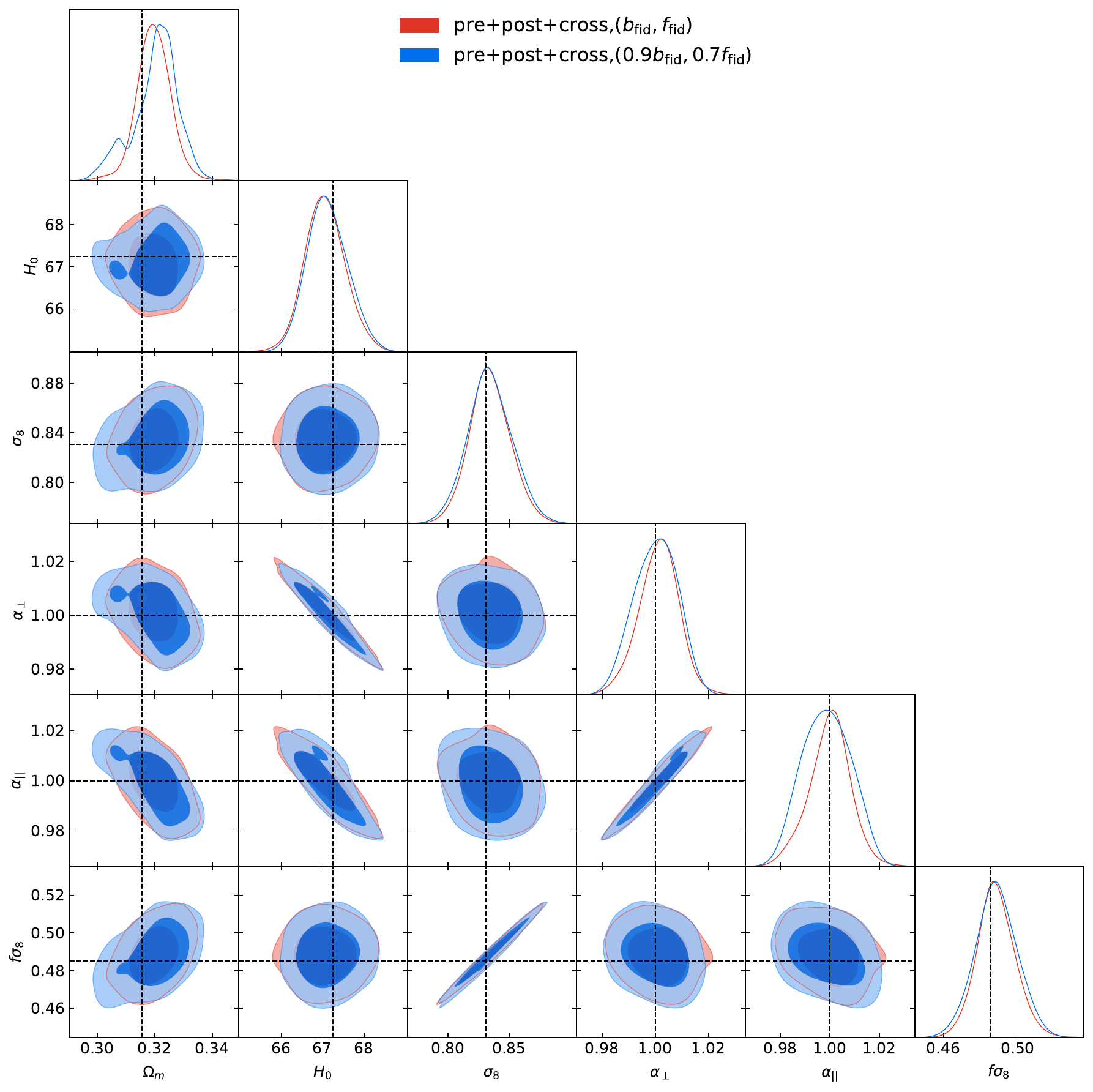}
\caption{The 1D posterior distribution and 2D contour plots showing 68\% and 95\% credible regions for the derived parameters ($\Omega_m$, $H_0$, $\sigma_8$, $\alpha_{\perp}$, $\alpha_{||}$, $f\sigma_8$) from $P_{\rm all}$ reconstructed using two different sets of $b_{\rm in}$ and $f_{\rm in}$ shown in the legend. The dashed lines show the expected values of the parameters.}
\label{fig:all_diff_bf}
\end{figure*}

\begin{figure*}[!t]
\centering
\includegraphics[scale=0.4]{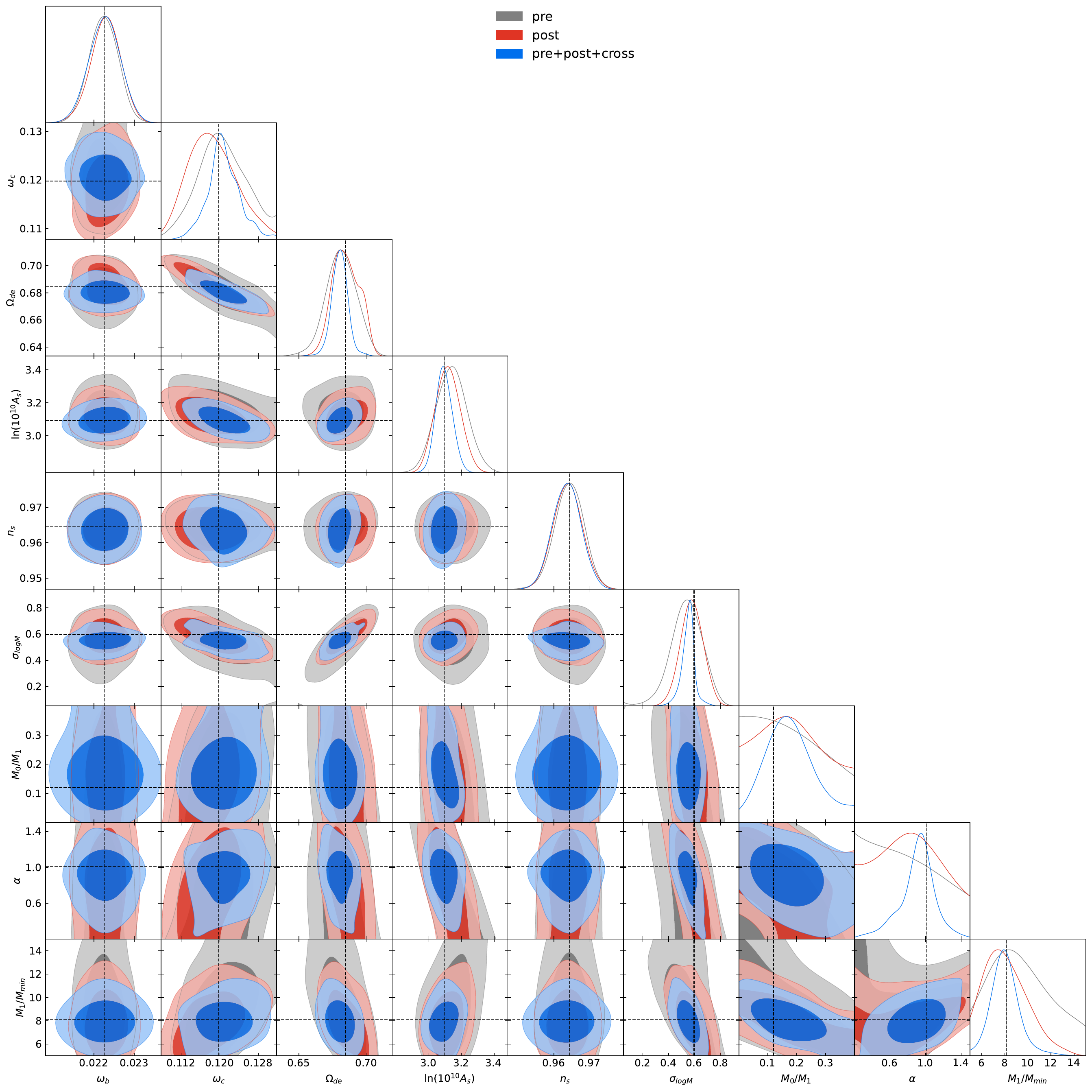}
\caption{The 1D posterior distribution and 2D contour plots showing 68\% and 95\% credible regions for cosmological and HOD parameters derived from combinations of different types of power spectra shown in the legend. The dashed lines show the expected values of the parameters.}
\label{fig:all}
\end{figure*}

\begin{figure*}[!t]
\centering
\includegraphics[scale=0.55]{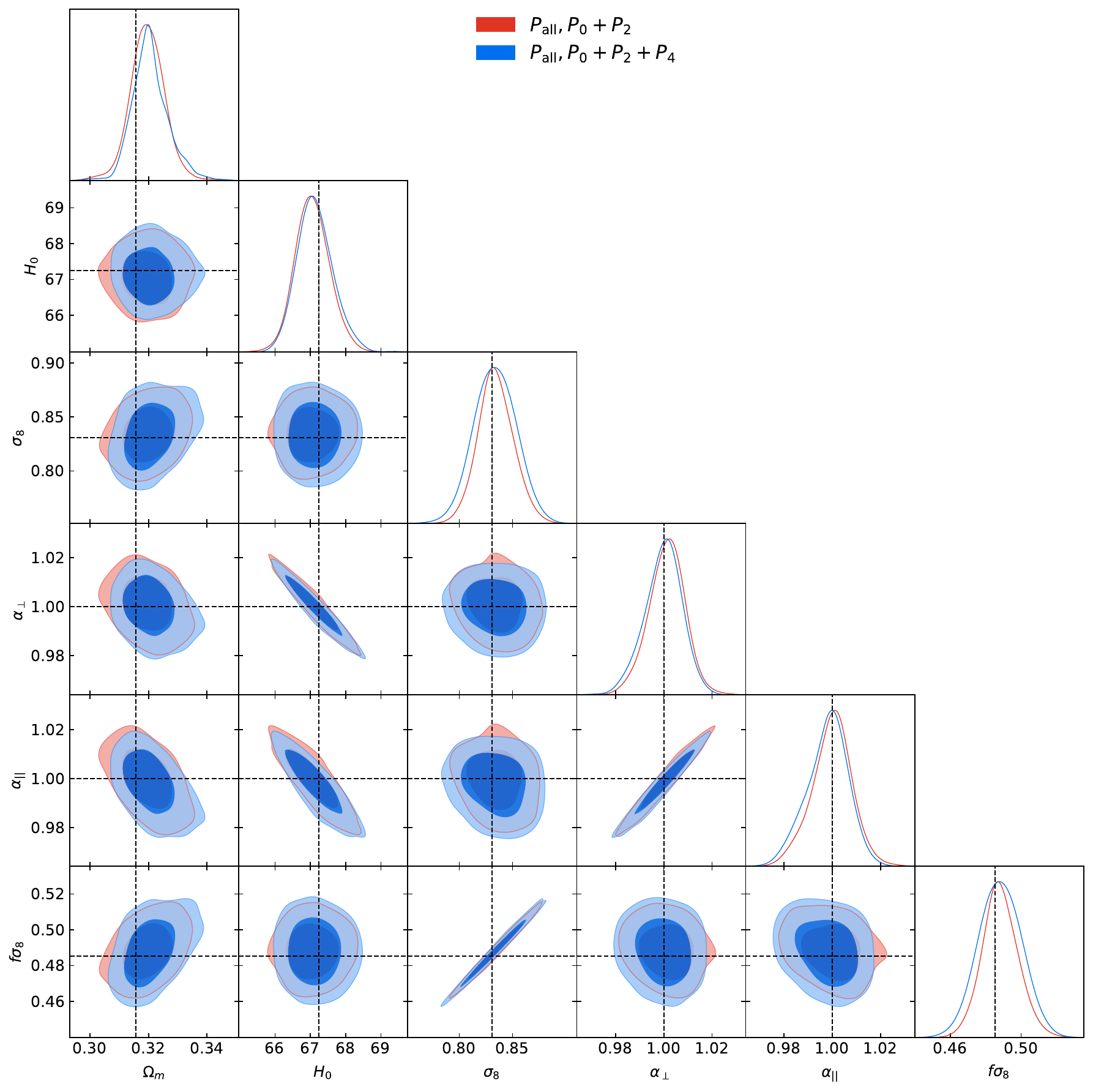}
\caption{The 1D posterior distribution and 2D contour plots showing 68\% and 95\% credible regions for ($\Omega_m$, $H_0$, $\sigma_8$, $\alpha_{\perp}$, $\alpha_{||}$, $f\sigma_8$) using $P_{\rm all}$ with and without the hexadecapole. The dashed lines show the expected values of the parameters.}
\label{fig:hexa}
\end{figure*}

\begin{figure*}[!t]
\centering
\includegraphics[scale=0.55]{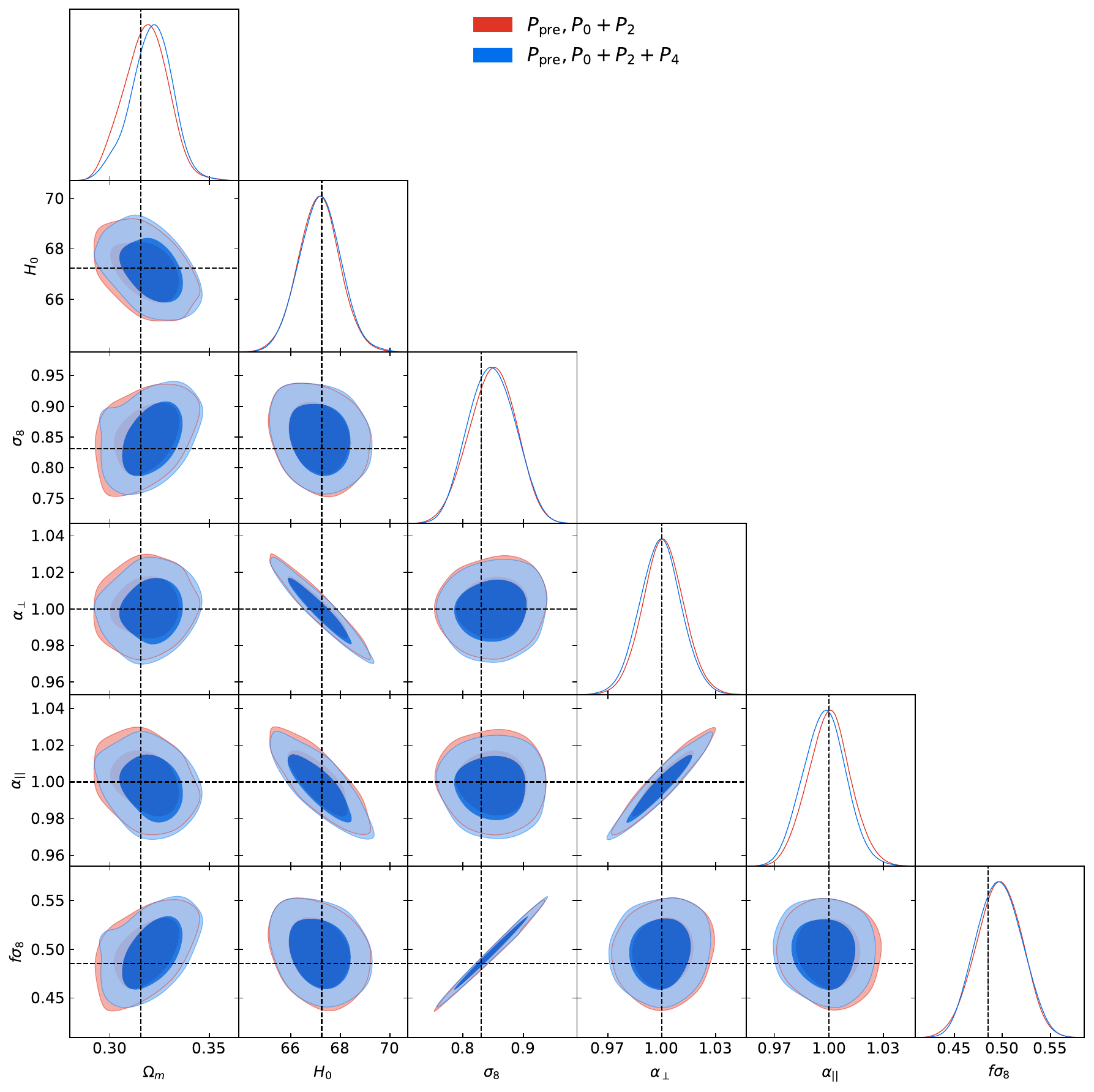}
\caption{Same as Fig. \ref{fig:hexa}, but using $P_{\rm pre}$ alone.}
\label{fig:prehexa}
\end{figure*}

\end{document}